\RequirePackage{pdfmanagement-testphase}
\DocumentMetadata{
  pdfstandard=A-2b,
  pdfversion=1.7,
  lang=en,
}
\documentclass[10pt,twoside]{book}

\usepackage{tagpdf}
\usepackage{adjustbox}
\usepackage{amsfonts}
\usepackage{amsmath}
\usepackage{amssymb}
\usepackage{array}
\usepackage{booktabs}
\usepackage{cite}
\usepackage{cuted}
\usepackage{etoolbox}
\usepackage{fancyhdr}
\usepackage[T1]{fontenc}
\usepackage[total={17cm,24cm}]{geometry}
\usepackage{imakeidx}
\usepackage[utf8]{inputenc}
\usepackage{listings}
\usepackage{longtable}
\usepackage{lscape}
\usepackage{mathtools}
\usepackage{multirow}
\usepackage{paralist}
\usepackage{rotating}
\usepackage{subcaption}
\usepackage{tabularray}
\usepackage{tcolorbox}
\usepackage[flushleft]{threeparttable}
\usepackage{times}
\usepackage[raggedright]{titlesec}
\usepackage{todonotes}
\usepackage[hyphens]{url}
\usepackage{verbatim}
\usepackage{wrapfig}
\usepackage{xcolor}
\usepackage[hang,flushmargin]{footmisc}
\usepackage{xparse}
\usepackage{expl3}
\usepackage{mfirstuc}
\usepackage{enumitem}
\usepackage{emptypage}
\usepackage{hyperxmp}
\usepackage[bookmarks=true,bookmarksopen=true,bookmarksopenlevel=0]{hyperref}
\hypersetup{
    pdftitle={An Ecosystem of Services for FAIR Computational Workflows},
    pdfauthor={Sean R. Wilkinson, Johan Gustafsson, Finn Bacall, Khalid Belhajjame, Salvador Capella, Jose Maria Fernandez Gonzalez, Jacob Fosso Tande, Luiz Gadelha, Daniel Garijo, Patricia Grubel, Bjorn Gruning, Farah Zaib Khan, Sehrish Kanwal, Simone Leo, Stuart Owen, Luca Pireddu, Line Pouchard, Laura Rodriguez-Navas, Beatriz Serrano-Solano, Stian Soiland-Reyes, Baiba Vilne, Alan Williams, Merridee Ann Wouters, Frederik Coppens, Carole Goble},
    pdfkeywords={scientific workflow systems; research data management; FAIR workflows; workflow repositories},
    pdflang={en},
    pdfcopyright={CC BY 4.0},
    pdflicenseurl={https://creativecommons.org/licenses/by/4.0/},
    hidelinks,
    pdfpagelayout=OneColumn,
    pdfdisplaydoctitle=true,
}

\makeindex[name=authors, columns=2, title=Authors, intoc]

\definecolor{chaptextbg}{RGB}{240, 240, 240}
\definecolor{chaptext}{RGB}{10, 10, 10}

\newtcolorbox{authorbox}{
  colback  = chaptextbg,
  colframe=  white!0,
  left=2mm,
  right=2mm,
  top=2mm,
  bottom=0mm,
  boxsep=0mm,
  arc=0mm,
}

\newcommand{\printtitlerun}[1]{
    \fancyhead[LO]{\hfill\MakeUppercase{\chaptertitlename} \thechapter. #1}
}

\newenvironment{abstract}{
    \begin{center}
        \begin{minipage}{0.85\textwidth}
            \noindent\textbf{%
                \color{chaptext}\MakeUppercase{Abstract: }%
            }}{\end{minipage}\end{center}}

\newcommand{\printbibliography}[1]{%
  \bibliographystyle{ieeetr}
  \bibliography{#1}
}

\titleformat{\chapter}[display]
  {\vspace{-50pt} \normalsize \large \color{chaptext}}%
  {%
    \color{chaptext}\MakeUppercase{\large\chaptertitlename}\hspace{1em}\thechapter%
  }%
  {0 pt}
  {\raggedright\LARGE\color{chaptext}\MakeUppercase}%

\definecolor{codegreen}{rgb}{0,0.6,0}
\definecolor{codegray}{rgb}{0.5,0.5,0.5}
\definecolor{codepurple}{rgb}{0.58,0,0.82}
\definecolor{backcolour}{rgb}{0.95,0.95,0.92}

\lstdefinestyle{mystyle}{
    backgroundcolor=\color{backcolour},   
    commentstyle=\color{codegreen},
    keywordstyle=\color{magenta},
    numberstyle=\tiny\color{codegray},
    stringstyle=\color{codepurple},
    basicstyle=\ttfamily\footnotesize,
    breakatwhitespace=false,         
    breaklines=true,                 
    captionpos=b,                    
    keepspaces=true,                 
    numbers=left,                    
    numbersep=5pt,                  
    showspaces=false,                
    showstringspaces=false,
    showtabs=false,                  
    tabsize=2
}

\colorlet{punct}{red!60!black}
\definecolor{background}{HTML}{EEEEEE}
\definecolor{delim}{RGB}{20,105,176}
\colorlet{numb}{magenta!60!black}

\lstdefinelanguage{json}{
    basicstyle=\normalfont\ttfamily,
    numbers=left,
    numberstyle=\scriptsize,
    stepnumber=1,
    numbersep=8pt,
    showstringspaces=false,
    breaklines=true,
    frame=lines,
    backgroundcolor=\color{background},
    literate=
     *{0}{{{\color{numb}0}}}{1}
      {1}{{{\color{numb}1}}}{1}
      {2}{{{\color{numb}2}}}{1}
      {3}{{{\color{numb}3}}}{1}
      {4}{{{\color{numb}4}}}{1}
      {5}{{{\color{numb}5}}}{1}
      {6}{{{\color{numb}6}}}{1}
      {7}{{{\color{numb}7}}}{1}
      {8}{{{\color{numb}8}}}{1}
      {9}{{{\color{numb}9}}}{1}
      {:}{{{\color{punct}{:}}}}{1}
      {,}{{{\color{punct}{,}}}}{1}
      {\{}{{{\color{delim}{\{}}}}{1}
      {\}}{{{\color{delim}{\}}}}}{1}
      {[}{{{\color{delim}{[}}}}{1}
      {]}{{{\color{delim}{]}}}}{1},
}

\makeatletter
\renewcommand\mainmatter{    \@mainmattertrue\cleardoublepage\renewcommand\thepage{\arabic{page}}}
\makeatother

\definecolor{backgroundColour}{HTML}{FAFAFA}
\definecolor{keywordclr}{HTML}{3F51B5}
\definecolor{commentclr}{HTML}{757575}
\definecolor{stringsclr}{HTML}{279049}
\definecolor{fnctionclr}{rgb}{0.467, 0, 0.533}
\definecolor{builtinclr}{rgb}{0.35, 0, 0.533}
\definecolor{symbolsclr}{rgb}{0.5, 0.25, 0.25}   
\definecolor{numbersclr}{rgb}{0.8, 0.2, 0}
\definecolor{bckgrndclr}{rgb}{0.91, 0.95, 0.95}
\lstdefinestyle{PythonStyle}{
    language=Python,
    backgroundcolor=\color{backgroundColour},
    keywordstyle=\color{keywordclr}\bfseries,
    stringstyle=\color{stringsclr},
    commentstyle=\color{commentclr}\itshape,
    upquote=true,
    basicstyle=\ttfamily\linespread{0.9}\scriptsize,
    breakatwhitespace=false,
    breaklines=true,
    captionpos=b,
    keepspaces=true,
    numbers=left,
    numbersep=5pt,
    numberstyle=\color{commentclr}\ttfamily\tiny,
    showspaces=false,
    showstringspaces=false,
    showtabs=false,
    tabsize=2,
    xleftmargin=1.25em,
    frame=single,
    framexleftmargin=1.25em,
    morekeywords={assert,with,as}
}
\usepackage{xparse}

\renewcommand{\footnotemargin}{1.3em}

\begin{document}

\pagestyle{fancy}
\fancyhead{}
\fancyfoot{}
\fancyhead[RE]{An Ecosystem of Services for FAIR Computational Workflows\hfill}
\fancyhead[LO]{\hfill\uppercase{\rightmark}}
\fancyfoot[LE]{}
\fancyfoot[C]{\thepage}
\fancyfoot[RO]{}

\mainmatter

\setcounter{chapter}{3}

\printtitlerun{An Ecosystem of Services for \MakeUppercase{FAIR} Computational Workflows}

\chapter{An Ecosystem of Services for \MakeUppercase{FAIR} Computational Workflows}\label{chap:50}
\begin{authorbox}
    \begin{minipage}{\textwidth}
      \footnotesize
      \textbf{Sean R. Wilkinson}\\
      Oak Ridge National Laboratory, Oak Ridge Leadership Computing Facility, Oak Ridge, TN, USA\\
      {\scriptsize\url{wilkinsonsr@ornl.gov}}\vspace{0.5em}
    \end{minipage}%
    
    \begin{minipage}{\textwidth}
      \footnotesize
      \textbf{Johan Gustafsson}\\
      University of Melbourne, Australian BioCommons, Melbourne, Australia\\
      {\scriptsize\url{johan@biocommons.org.au}}\vspace{0.5em}
    \end{minipage}%
    
    \begin{minipage}{\textwidth}
      \footnotesize
      \textbf{Finn Bacall}\\
      The University of Manchester, Manchester, UK\\
      {\scriptsize\url{finn.bacall@manchester.ac.uk}}\vspace{0.5em}
    \end{minipage}%
    
    \begin{minipage}{\textwidth}
      \footnotesize
      \textbf{Khalid Belhajjame}\\
      University Paris-Dauphine, France\\
      {\scriptsize\url{khalid.belhajjame@dauphine.fr}}\vspace{0.5em}
    \end{minipage}%
    
    \begin{minipage}{\textwidth}
      \footnotesize
      \textbf{Salvador Capella}\\
      Barcelona Supercomputing Center, Barcelona, Spain\\
      {\scriptsize\url{salvador.capella@bsc.es}}\vspace{0.5em}
    \end{minipage}%
    
    \begin{minipage}{\textwidth}
      \footnotesize
      \textbf{Jose Maria Fernandez Gonzalez}\\
      Barcelona Supercomputing Center, Barcelona, Spain\\
      {\scriptsize\url{jose.m.fernandez@bsc.es}}\vspace{0.5em}
    \end{minipage}%

    \begin{minipage}{\textwidth}
      \footnotesize
      \textbf{Jacob Fosso Tande}\\
      North Carolina State University, Raleigh, NC, USA\\
      {\scriptsize\url{jfossot@ncsu.edu}}\vspace{0.5em}
    \end{minipage}%
    
    \begin{minipage}{\textwidth}
      \footnotesize
      \textbf{Luiz Gadelha}\\
      German Cancer Research Center, Heidelberg, Germany\\
      {\scriptsize\url{luiz.gadelha@dkfz-heidelberg.de}}\vspace{0.5em}
    \end{minipage}%
    
    \begin{minipage}{\textwidth}
      \footnotesize
      \textbf{Daniel Garijo}\\
      Universidad Polit\'{e}cnica de Madrid, Department, City, Country\\
      {\scriptsize\url{daniel.garijo@upm.es}}\vspace{0.5em}
    \end{minipage}%
    
    \begin{minipage}{\textwidth}
      \footnotesize
      \textbf{Patricia Grubel}\\
      Los Alamos National Laboratory, Los Alamos, NM, USA\\
      {\scriptsize\url{pagrubel@lanl.gov}}\vspace{0.5em}
    \end{minipage}%
    
    \begin{minipage}{\textwidth}
      \footnotesize
      \textbf{Bj\"{o}rn Grüning}\\
      University of Freiburg, Freiburg im Breisgau, Germany\\
      {\scriptsize\url{gruening@informatik.uni-freiburg.de}}\vspace{0.5em}
    \end{minipage}%
    
    \begin{minipage}{\textwidth}
      \footnotesize
      \textbf{Farah Zaib Khan}\\
      University of Melbourne, Australian BioCommons, Melbourne, Australia\\
      {\scriptsize\url{khanf1@unimelb.edu.au}}\vspace{0.5em}
    \end{minipage}%
    
    \begin{minipage}{\textwidth}
      \footnotesize
      \textbf{Sehrish Kanwal}\\
      University of Melbourne, Collaborative Centre for Genomics Cancer Medicine, Melbourne, Australia\\
      {\scriptsize\url{kanwals@unimelb.edu.au}}\vspace{0.5em}
    \end{minipage}%
    
    \begin{minipage}{\textwidth}
      \footnotesize
      \textbf{Simone Leo}\\
      CRS4, Sardinia, Italy\\
      {\scriptsize\url{simone.leo@crs4.it}}\vspace{0.5em}
    \end{minipage}%

\end{authorbox}

\begin{authorbox}

    \begin{minipage}{\textwidth}
      \footnotesize
      \textbf{Stuart Owen}\\
      The University of Manchester, Manchester, UK\\
      {\scriptsize\url{sowen@cs.manchester.ac.uk}}\vspace{0.5em}
    \end{minipage}%
    
    \begin{minipage}{\textwidth}
      \footnotesize
      \textbf{Luca Pireddu}\\
      CRS4, Sardinia, Italy\\
      {\scriptsize\url{pireddu@crs4.it}}\vspace{0.5em}
    \end{minipage}%
    
    \begin{minipage}{\textwidth}
      \footnotesize
      \textbf{Line Pouchard}\\
      Sandia National Laboratories, Albuquerque, NM, USA\\
      {\scriptsize\url{lcpouch@sandia.gov}}\vspace{0.5em}
    \end{minipage}%
    
    \begin{minipage}{\textwidth}
      \footnotesize
      \textbf{Laura Rodr\'{i}guez-Navas}\\
      Universitat de Vic, Barcelona, Spain\\
      {\scriptsize\url{laura.rodriguez.navas@uvic.cat}}\vspace{0.5em}
    \end{minipage}%
    
    \begin{minipage}{\textwidth}
      \footnotesize
      \textbf{Beatriz Serrano-Solano}\\
      Euro-Bioimaging, Germany\\
      {\scriptsize\url{beatriz.serrano.solano@eurobioimaging.eu}}\vspace{0.5em}
    \end{minipage}%
    
    \begin{minipage}{\textwidth}
      \footnotesize
      \textbf{Stian Soiland-Reyes}\\
      The University of Manchester, Manchester, UK\\
      {\scriptsize\url{soiland-reyes@manchester.ac.uk}}\vspace{0.5em}
    \end{minipage}%
    
    \begin{minipage}{\textwidth}
      \footnotesize
      \textbf{Baiba Vilne}\\
      Riga Stradi\textcommabelow{n}\v{s} University, Bioinformatics Group, Latvia\\
      {\scriptsize\url{baiba.vilne@rsu.lv}}\vspace{0.5em}
    \end{minipage}%
    
    \begin{minipage}{\textwidth}
      \footnotesize
      \textbf{Alan Williams}\\
      The University of Manchester, Manchester, UK\\
      {\scriptsize\url{alan.r.williams@manchester.ac.uk}}\vspace{0.5em}
    \end{minipage}%
    
    \begin{minipage}{\textwidth}
      \footnotesize
      \textbf{Merridee Ann Wouters}\\
      University of New South Wales, Sydney, Australia\\
      {\scriptsize\url{merridee.wouters@unsw.edu.au}}\vspace{0.5em}
    \end{minipage}%
    
    \begin{minipage}{\textwidth}
      \footnotesize
      \textbf{Frederik Coppens}\\
      Vlaams Instituut voor Biotechnologie, Belgium\\
      {\scriptsize\url{frederik.coppens@vib.be}}\vspace{0.5em}
    \end{minipage}%
    
    \begin{minipage}{\textwidth}
      \footnotesize
      \textbf{Carole Goble}\\
      The University of Manchester, Manchester, UK\\
      {\scriptsize\url{carole.goble@manchester.ac.uk}}\vspace{0.5em}
    \end{minipage}%
\end{authorbox}

\tiny 

\normalsize

\begin{abstract}
Computational workflows represent major investments of effort and expertise. As first-class, publishable research objects of their own, they are key to sharing methodological know-how for reuse, reproducibility, and transparency. Thus, the application of the \MakeUppercase{FAIR} Principles to workflows is inevitable to enable them to be Findable, Accessible, Interoperable, and Reusable. Making workflows \MakeUppercase{FAIR} reduces duplication of effort, assists in the reuse of best practice approaches and community-supported standards, and ensures that workflows as digital objects can support reproducible, robust science. \MakeUppercase{FAIR} workflows draw from both \MakeUppercase{FAIR} data and software principles, and they help ensure and support data FAIRification.

The \MakeUppercase{FAIR} Principles emphasize the association of persistent identifiers and machine-actionable metadata with workflows. Implementing the Principles requires a framework with appropriate programmatic protocols and an accompanying ecosystem of services, tools, policies, and best practices, as well the buy-in of existing workflow systems. The European EOSC-Life Workflow Collaboratory is an example of such a digital infrastructure for the Biosciences. It includes a metadata standards framework for describing workflows that is managed and used by dedicated new \MakeUppercase{FAIR} workflow services and programmatic APIs for interoperability and metadata access. It includes the WorkflowHub registry and LifeMonitor workflow testing service, and it incorporates existing workflow systems and packaging solutions.

Here, we introduce the \MakeUppercase{FAIR} Principles for workflows and connect \MakeUppercase{FAIR} workflows with the \MakeUppercase{FAIR} ecosystems they inhabit with the EOSC-Life Collaboratory as a concrete example. We also introduce other community efforts that are easing the ways that workflows are shared and reused by others, and we discuss how the variations in different workflow settings impact their \MakeUppercase{FAIR} perspectives.

\end{abstract}

\section{Introduction}

Computational workflows represent major investments of both effort and expertise. They are first class, publishable research objects in their own right and key to sharing methodological know-how for reuse, reproducibility and transparency, as well as being reusable as executable software, given appropriate portability between or access to execution environments.

The \MakeUppercase{FAIR} (Findable, Accessible, Interoperable, Reusable) principles have laid a foundation for sharing and publishing digital assets, starting with data \cite{wilkinson_2016}. At their heart, the principles set out to promote the widespread use of persistent identifiers (PID) and machine-actionable metadata, along with the use of programmatic protocols to access objects and their metadata. They also propose guidelines for: the form and content that the metadata should take; restricted and long term access to the metadata; services for findability (registries and search) and accessibility (authentication and authorization); and clear licenses that explain the conditions under which others are permitted to use and modify the data. The principles have had a remarkable effect on the scientific, public sector, and even commercial community, and have almost become doctrine in the past decade. 
\MakeUppercase{FAIR} principles have been extended to all digital objects, including software \cite{barker_2022} and specialised forms of software such as Machine Learning Models \cite{huerta_2023}. Inevitably, they have also been extended to computational workflows \cite{goble_2019,wilkinson_2025}. The \MakeUppercase{FAIR} data principles themselves originate from a desire to support automated data processing, by emphasizing machine accessibility of data and metadata. Workflows are special kinds of software, but they're also a precise description of a process. As workflows are digital objects that have a dual role as software and explicit method description, their \MakeUppercase{FAIR} properties draw from both data and software principles.

Making workflows \MakeUppercase{FAIR} should reduce duplication of effort and assist in the reuse of best practice approaches and community-supported standards. \MakeUppercase{FAIR} workflows should also encourage interdisciplinary collaboration, enabling workflows developed in one field to be repurposed and adapted for use in other research domains. Implementing \MakeUppercase{FAIR} principles in computational workflow is a step to addressing the growing need for efficient, transparent, and reproducible research practices. By making workflows FAIR, researchers can maximise the value and impact of their work, accelerate scientific discovery, and promote collaboration and knowledge sharing. Workflows also aid in ensuring that the data products used and produced are FAIR. 

At the heart of \MakeUppercase{FAIR} is machine-actionable metadata, driving the use of standardised formats and metadata schemas as well as persistent identifiers for discovery, access and retrieval. A documented workflow promotes interoperability. A documented provenance record of the computational process makes it easier to understand and rerun, ensuring it can be replicated and results validated, thus supporting team collaboration as well as standardizing processes and analysis methods. Better metadata and identifiers for executed codes/data simplify the process of modifying and adapting data analyses to incorporate additional and/or different methods, tools, or data sources.

The implementation of \MakeUppercase{FAIR} principles for workflows requires:
\begin{itemize}
  \item a persistent identifier and metadata framework, with appropriate programmatic protocols,
  \item an ecosystem of interoperable tools and services, fully describing the context necessary for discovering, accessing and executing  workflows, and
  \item a community of stakeholders with the capacity and capability to make, (re)use, and manage workflows and workflow services, and curate workflows.
\end{itemize}

The European EOSC-Life\footnote{\url{https://www.eosc-life.eu/}} Workflow Collaboratory is an example of such a digital infrastructure for the Biosciences. A metadata standards framework for describing workflows, including RO-Crate, Bioschemas, and CWL is managed and used by dedicated, new \MakeUppercase{FAIR} workflow services. This initiative developed a framework, ecosystem of tools and a stakeholder community for researchers and developers in the Life Science community, contributing to the European Open Science Cloud\footnote{See \url{https://eosc.eu/eosc-about/}.}.

The WorkflowHub registry supports workflow Findability and Accessibility, while workflow testing services like LifeMonitor\footnote{See \url{https://crs4.github.io/life_monitor/}.} support long-term Reusability, Usability and Reproducibility (a specialised form of Reuse). Existing workflow management systems/languages (e.g.,, Nextflow\footnote{\url{https://www.nextflow.io/}}~\cite{ditommaso_2017}, Galaxy\footnote{\url{https://galaxyproject.org/}}~\cite{abueg_2024}, Snakemake\footnote{\url{https://snakemake.github.io/}}~\cite{moelder2021}, Common Workflow Language CWL\footnote{\url{https://www.commonwl.org/}}~\cite{crusoe_2022}, Workflow Description Language WDL\footnote{\url{https://openwdl.org/}}~\cite{voss_2017}) and packaging solutions (e.g., Conda\footnote{\url{https://anaconda.org/}}, Docker\footnote{\url{https://www.docker.com/}}, Apptainer\footnote{\url{https://apptainer.org/} (formerly Singularity)}) are incorporated and adapted to promote portability, composability, interoperability, provenance collection and reusability, and to use and support these \MakeUppercase{FAIR} services. 

The Collaboratory approach and services has been adopted by ELIXIR\footnote{\url{https://elixir-europe.org }}, the European Research Infrastructure for Life Science Data and the Australian BioCommons\footnote{\url{https://www.biocommons.org.au/}}~\cite{francis_2023}, who work together to continue to develop and manage the infrastructure and promote its use to stakeholders. Combinations of services have been adopted by many other Life Science consortia and projects, and initiatives outside of Life Sciences.

In this chapter, we will introduce the \MakeUppercase{FAIR} Principles for workflows and the background of the EOSC-Life Workflow Collaboratory. We then use the Collaboratory as a concrete example to make the connections between \MakeUppercase{FAIR} workflows and the \MakeUppercase{FAIR} ecosystems in which they live, and sketch the workflow's metadata journey through its life cycle and use of the services. We will also introduce other community efforts that are easing the ways that workflows are shared and reused by others, and discuss how the variations in different workflow settings impact their \MakeUppercase{FAIR} perspective.

\section{\MakeUppercase{FAIR} Principles Applied to Workflows}

Workflows defy simple definitions \cite{ferreiradasilva_2022}. For our purposes, we consider a workflow as a specialised piece of software (to which \MakeUppercase{FAIR} applies) which has components which may be individually \MakeUppercase{FAIR} and which produces \MakeUppercase{FAIR} data products by design. We do not assume that a \MakeUppercase{FAIR} workflow must require \MakeUppercase{FAIR} input data and parameters, but we do believe a \MakeUppercase{FAIR} workflow should produce \MakeUppercase{FAIR} data products such as \MakeUppercase{FAIR} output data and \MakeUppercase{FAIR} workflow run data.

\MakeUppercase{FAIR} Workflows draw from both \MakeUppercase{FAIR} data and software principles. Workflows propose explicit method abstractions and tight bindings to data, hence making many of the data principles apply. Meanwhile, as executable pipelines with a strong emphasis on code composition and data flow between steps, the software principles apply, too. As workflows are chiefly concerned with the processing and creation of data, they also have an important role to play in ensuring data is FAIR. 

Although workflows are inherently software, workflow management systems (WMS) have additional properties that impact how we might apply \MakeUppercase{FAIR} principles \cite{suter_2025}. Applying \MakeUppercase{FAIR} to workflows involves considering workflows as both data and software descriptions of a process, and then building on the existing principles. This means addressing both the executable components and the associated metadata that describes the workflow's purpose, inputs, outputs, dependencies, and execution environment.

\textbf{\textit{Method abstraction.}} Workflows have a specification, which is a description of the steps with parameters and inputs and guidance -- offering \MakeUppercase{FAIR} transparency, and metadata descriptions. We can consider these almost to be like \MakeUppercase{FAIR} data because they're descriptive artifacts. On the other hand, we have software, including WMS and any tools and infrastructure required by individual codes for the orchestration of the workflow. Like all software, this is related to reproducibility, being able to run those pipelines and reuse those pipelines. The descriptions refer to method preservation. Alongside these two perspectives of method abstraction and the software that implements the method are the associated objects around workflows: logs of their execution, example data, test data, and services associated with them in order to be able to check whether these workflows are FAIR.

\textbf{\textit{Method modularization and composability.}} Workflows expect to take various different components in different languages from different third parties, compose them, and port them to different hosts. Workflows are compositions of components, including other workflows that can be broken down, versioned, recycled and so on. This requires \MakeUppercase{FAIR} to apply at the different levels of abstraction at the description level and the software level, and for the different components that make up the workflows. There are multiple workflow systems in the landscape, which typically are used in an intertwined kind of way. People use a workflow management system, which is a dedicated infrastructure that preserves a neat "separation of concerns" with respect to modularization, abstraction, and execution. Executed components are also typically used with interactive notebooks, and scripting environments, which still run multiple steps but are less clean in their separation of concerns.

\textbf{\textit{Enhancing interoperability and automation.}} Beyond software reproducibility, maintaining a consistent computational environment is crucial for \MakeUppercase{FAIR} workflows. Interoperability remains a significant challenge across research domains and groups. \MakeUppercase{FAIR} workflows require an ecosystem of interoperable tools and services, fully describing the context necessary for executing the workflows as software, for example Infrastructure as Code tools that allow researchers to define, provision, and manage computational infrastructure in a declarative manner using configuration files, ensuring that workflow execution environments remain consistent across on-premises HPC clusters and different cloud platforms.

\textbf{\textit{Instruments of data generation.}} Because workflows typically deal with data flow, they should be supporting \MakeUppercase{FAIR} data. Thus, we need to test that the data generated by workflows are actually FAIR. For example, we can test that workflows are licensing data outputs, using community data formats, noting required usage restrictions, and handling identifiers correctly, the last of which is critical for the detailed provenance of the data handled by those workflows. Because \MakeUppercase{FAIR} is about machine-actionable metadata, such tests lend themselves well to automation, which means workflows can be used to test the FAIRness of the data produced by other workflows, too.

\textbf{The Workflow Community Initiative \MakeUppercase{FAIR} Workflows Working Group (WCI-FW)} systematically addressed the application of both \MakeUppercase{FAIR} data and software principles to computational workflows \cite{wilkinson_2025}:
\begin{itemize}
    \item \textbf{Findable:} Workflows and their components are easily discoverable through rich metadata, unique and persistent identifiers (e.g., DOIs), and registration in searchable repositories.
    \item \textbf{Accessible:} Workflows can be retrieved and accessed using standard protocols, with clear guidelines on access rights and authentication if needed.
    \item \textbf{Interoperable:} Workflows use standardised formats, controlled vocabularies, and metadata schemas to ensure compatibility and integration with other resources across infrastructures.
    \item \textbf{Reusable:} Workflows are well-documented, licensed, and versioned, with clear information on dependencies and execution requirements, enabling others to easily adapt and reuse them for their research.
\end{itemize}
Detailed metadata and documentation include descriptive metadata such as title, authors, creation date, and version, as well as version histories when possible. The FAIRification process involves the use of persistent identifiers, accessible data retrieval methods and clear licenses that explain the conditions under which others are permitted to use and modify parts or the whole workflow. 

\begin{table}[ht]\tiny
\caption{The \MakeUppercase{FAIR} Principles Applied to Computational Workflows \cite{wilkinson_2025}.}\label{tab:1}
\begin{tabular}{|p{0.6\textwidth}|p{0.25\textwidth}|}
\hline
Guideline & Corresponding [D]ata~\cite{wilkinson_2016} and [S]oftware~\cite{barker_2022} principles \\
\hline
F1. A workflow is assigned a globally unique and persistent identifier.
                                            & D-F1 and S-F1\\ \hline & \\
   F1.1. Components of the workflow representing levels of granularity are assigned distinct identifiers.& S-F1.1 \\ \hline & \\
F1.2. Different versions of the workflow are assigned distinct identifiers.
                                                        & S-F1.2 \\ \hline & \\
F2. A workflow and its components are described with rich metadata.
                                            & D-F2 and S-F2 \\ \hline & \\
F3. Metadata clearly and explicitly include the identifier of the workflow, and workflow versions, that they describe.
                                            & D-F3  and  S-F3 \\ \hline & \\
F4. Metadata and workflow are registered or indexed in a searchable \MakeUppercase{FAIR} resource.
                                            & D-F4  and  S-F4 \\ \hline & \\
A1. Workflow and its components are retrievable by their identifiers using a standardised communications protocol.
                                            & D-A1  and  S-A1 \\ \hline & \\
A1.1. The protocol is open, free, and universally implementable.
                                        & D-A1.1  and  S-A1.1 \\ \hline & \\
A1.2. The protocol allows for an authentication and authorization procedure, when necessary.
                                        & D-A1.2  and  S-A1.2 \\ \hline & \\
A2. Metadata are accessible, even when the workflow is no longer available.
                                        & D-A2  and  S-A2 \\ \hline & \\
I1. Workflow and its metadata (including workflow run provenance) use a formal, accessible, shared, transparent, and broadly applicable language for knowledge representation.
                                        & D-I1  and  S-R1.2 \\ \hline & \\
I2. Metadata and workflow use vocabularies that follow \MakeUppercase{FAIR} principles.
                                        & D-I2 \\ \hline & \\
I3. Workflow is specified in a way that allows its components to read, write, and exchange data (including intermediate data), in a way that meets domain-relevant standards.
                                        & D-R3  and  S-I1 \\ \hline & \\
I4. Workflow and its metadata (including workflow run provenance) include qualified references to other objects and the workflow's components.
                            & D-I3, S-I2,  and  S-R1.2 \\ \hline & \\
R1. Workflow is described with a plurality of accurate and relevant attributes.
                                & D-R1  and  S-R1 \\ \hline & \\
R1.1. Workflow is released with a clear and accessible license.
                                & D-R1.1  and  S-R1.1 \\ \hline & \\
R1.2. Components of the workflow representing levels of granularity are given clear and accessible licenses.
                                & D-R1.1  and  S-R1.1 \\ \hline & \\
R1.3. Workflow is associated with detailed provenance of the workflow and of the products of the workflow.
                                & D-R1.2  and  S-R1.2 \\ \hline & \\
R2. Workflow includes qualified references to other workflows.
                                & D-I3  and  S-R2 \\ \hline & \\
R3. Workflow meets domain-relevant community standards.
                                & D-R1.3  and  S-R3 \\ \hline
\end{tabular}
\footnotetext[1]{The "Based on" column references the source rule from the \MakeUppercase{FAIR} Principles for [D]ata~\cite{wilkinson_2016} and [S]oftware~\cite{barker_2022}.}
\end{table}

\section{The \MakeUppercase{EOSC-L}ife \MakeUppercase{FAIR} Workflow Collaboratory}

The EOSC-Life project brought together Europe's Life Science Research Infrastructures, ranging from biobanking and clinical trials to plant phenotyping and bioimaging, to create a FAIR, digital and collaborative space for biological and medical research. The Life Science infrastructures extensively use computational workflows for preparing, analysing, and increasingly sharing large volumes of data as well as simulations and predictive models. 

The community is large and diverse. The heterogeneity of the Research Infrastructures is reflected in the diversity of their data analysis practices and the variety of WMS they use, including specialist platforms. Each has different kinds of WMS used for building multi-step pipelines and multi-step processes to coordinate and execute multiple codes, often codes that they may not have developed themselves. The workflow systems are handling data and processing dependencies and doing other kinds of heavy lifting, typically data pipelines but also simulations, and predictions. Some WMS and languages, like Galaxy \cite{abueg_2024}, are primarily designed for end-users to assemble toolchains  running on a shared managed community instance over HPC, Cloud, and cluster infrastructure. Others, like Snakemake \cite{koester_2012}, are Python-based pipelines used by expert bioinformaticians for personal exploratory analysis that sometimes move to local production pipelines.  Outside of a handful of widely used WMS in Biosciences \cite{reiter_2021} (Galaxy, Nextflow, Snakemake, CWL, WDL), there are many specialist software frameworks such as Scipion 3\footnote{\url{https://scipion.i2pc.es/}}, which is used in structural biology, specifically for cryo-electron microscopy image processing. 

\begin{figure}[htbp!]
\centering
\includegraphics[width=.9\textwidth]{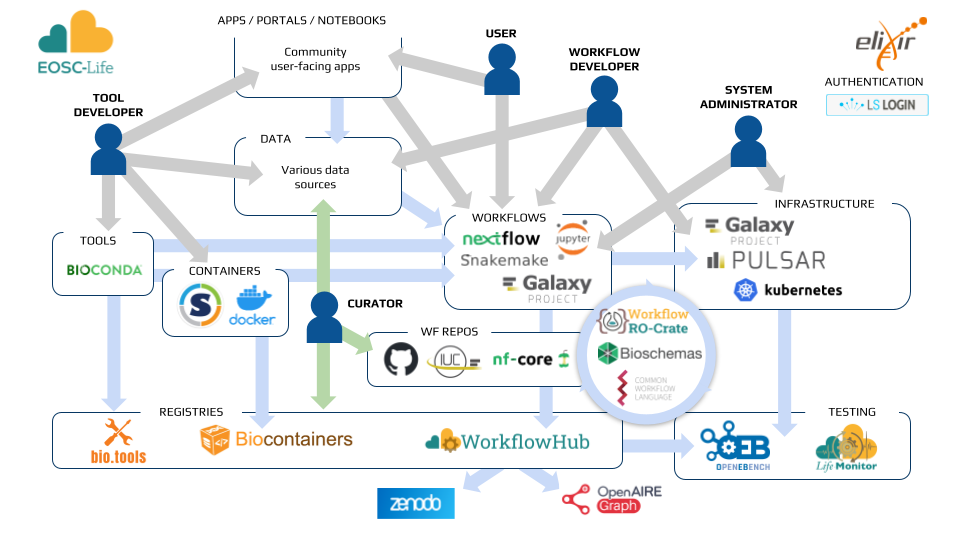}
\caption{\label{fig:1}EOSC-Life \MakeUppercase{FAIR} Workflow Collaboratory.}
\end{figure}

To serve this broad-ranging community, the Workflow Collaboratory honors this diversity and legacy and developed (i) a PID and metadata framework and (ii) an ecosystem of services for researchers and workflow specialists to find, use and reuse workflows, and deploy them in Cloud, HPC or on-premise. As shown in Figure~\ref{fig:1}, there are many WMS and languages in use there alongside dedicated registries and repositories that work with dedicated workflow services.

The Collaboratory framework and services are now coordinated  by ELIXIR, the European Research Infrastructure for Life Science Data, the Australian BioCommons, working with other large bioscience communities such as Biobanking (BBMRI-ERIC), biodiversity (DISSCo, LifeWatch) and communities outside of biology such as Climate change (ClimateAdapt4EOSC). The EuroScienceGateway\footnote{\url{https://galaxyproject.org/projects/esg/}} project notably supported the development of the services, including deployment on the European Open Science Cloud and HPC platforms,  and widened their takeup outside Bioinformatics in astronomy, climate change, particle physics and social sciences.

Projects that have used the Collaboratory's metadata framework and services and contribute to its development include:
\begin{compactitem}
    \item \textbf{Biodiversity Genomics Europe}\footnote{\url{https://biodiversitygenomics.eu/}, \url{https://doi.org/10.3030/101059492}} uniting BIOSCAN Europe pipelines for DNA barcoding, and the European Reference Genome Atlas (ERGA), pipelines for genome sequencing, to support biodiversity conservation across Europe. 
    \item \textbf{Biodiversity Digital Twin (BioDT)}\footnote{\url{https://biodt.eu/}} developing a series of workflows aimed at enhancing understanding of biodiversity dynamics through advanced modeling and simulation.
    \item \textbf{eFlows4HPC}\footnote{\url{https://eflows4hpc.eu/}} focused on High-Performance Computing simulations with data analytics and machine learning, facilitating the creation and execution of complex applications across various domains.
    \item \textbf{BioExcel}\footnote{\url{https://bioexcel.eu/}} Center of Excellence for Computational Biomolecular Research whose Building Blocks (BioBB) library offers a suite of pre-configured workflows designed to streamline various biomolecular simulation tasks. 
    \item \textbf{MGNify}\footnote{\url{https://www.ebi.ac.uk/metagenomics/pipelines/}} specialised analysis pipelines tailored for different types of microbiome sequence data. 
    \item \textbf{Australian BioCommons} suite of bioinformatics workflows designed to support life science researchers in processing and analyzing complex biological data. 
\end{compactitem}

To implement this \MakeUppercase{FAIR} Workflows Collaboratory we address the following four areas: 
\begin{compactenum}
    \item A metadata, PID and protocols standards framework for workflows;\
    \item An ecosystem of interoperable services using 1;
    \item Support for the lifecycle of the workflow, and its metadata, by the services of 1 and 2 to workflows  \MakeUppercase{FAIR} by design and make their data products \MakeUppercase{FAIR} by design;
    \item Engaging and onboarding the stakeholders responsible for using and delivering \MakeUppercase{FAIR} workflows and their services.
\end{compactenum}
Each of these will be discussed in the following sections.

\section{\MakeUppercase{FAIR} Metadata Framework and Service Ecosystem}

The foundational component of the Collaboratory is a standards framework that enables a common understanding of workflows and their \MakeUppercase{FAIR} requirements. This includes their persistent identifiers and metadata descriptors; standards for  workflows, associated data and their metadata reporting, citation, publishing and exchange between services and repositories; and standard APIs.

\subsection{Metadata Framework}

A common understanding of a given workflow requires standards for the metadata that describes that workflow, including its purpose, software and other workflow dependencies and associated companion components (e.g., reference data, test data). The Collaboratory uses three major components for its metadata framework:
\begin{compactitem}
    \item \textit{A collection of metadata schemas and ontologies} for annotating the workflow, tools and parameters for findability and reusability. This includes crosswalks to other software metadata schema (such as CodeMeta\footnote{\url{https://codemeta.github.io/index.html}}) for interoperability and feeds community scholarship knowledge graphs (such as DataCite's PID Graph\footnote{\url{https://datacite.org/blog/introducing-the-pid-graph/}} and openAIRE's Scholarly Knowledge Graph\footnote{\url{https://graph.openaire.eu/}}) for findability;
    \item \textit{A workflow description language independent of execution language and platforms} for canonically describing the steps of the workflow, improving portability and interoperability;
    \item \textit{A \MakeUppercase{FAIR} Digital Object format} for describing and packaging workflows, metadata, associated companion data objects, execution provence log, for exchange, reporting  archiving and citation, and for supporting findability, accessibility, interoperability and reusability.
\end{compactitem}

\subsubsection{Metadata schema and ontologies}

\MakeUppercase{FAIR} implementations refer to the use of "Semantic Artefacts" \cite{corcho_2024} used for describing the metadata associated with objects, in our case workflows and their companion tools/code and data. Semantic Artefacts, broadly, come in two forms:
\begin{compactitem}
    \item \textit{standardise the description} using schemas (also known as information  models, reporting guides, formats) with attribute-value pairs and indications which are mandatory, recommended or optional; 
    \item \textit{define what we mean} using common, shared terms/concepts (also known as terminology, controlled vocabulary, ontology,  thesaurus, dictionary, tag library, taxonomy, code list etc) for the attributes and values. 
\end{compactitem}
For our framework we use schema.org, Linked Data (JSON-LD) and community ontologies. Schema.org is a community activity founded by  commercial search engines to create structured data on the web. Its vocabularies cover entities, relationships between entities and actions, and can be extended through a well-documented extension model. The Bioschemas initiative \cite{gray_2017} cherry-picks from the schema.org vocabularies to make opinionated profiles. 

To \textit{standardise the descriptions} we developed three Bioschemas profiles: ComputationalTool\footnote{\url{https://bioschemas.org/profiles/ComputationalTool/1.0-RELEASE}}, ComputationalWorkflow\footnote{\url{https://bioschemas.org/profiles/ComputationalWorkflow/1.0-RELEASE}} and FormalParameter\footnote{\url{https://bioschemas.org/profiles/FormalParameter/1.0-RELEASE}} (an identified variable used to stand for the actual value(s) consumed/produced by a ComputationalWorkflow). The profiles cover the minimum, recommended, and optional levels of metadata completeness for both tools (i.e. software) and workflows. Table~\ref{tab:2} below shows an example alignment between a selection of metadata fields from both standards, and their level. These profiles are agnostic to discipline and provide a standard for addressing the requirements of the \MakeUppercase{FAIR} software and workflow principles: this includes identifiers for the digital objects (F1), their levels of granularity (F1.1), and versions (F1.2), rich metadata that features identifiers and versions (F2/F3), the provision of a standard for knowledge representation (workflows-I1 / software-R1.2), as well as including accurate and relevant attributes (R1), licenses (workflows-R1.1 / software-1.2), provenance (workflows-R1.3 / software-R1.2), and references to other workflows/software (R2).

\begin{table}[ht]\footnotesize
\caption{Bioschemas profile comparison for a selection of computational tool and computational workflow metadata fields.}\label{tab:2}
\begin{tabular}{|p{0.3\textwidth}|p{0.25\textwidth}|p{0.3\textwidth}|}
\hline
Metadata field & Computational Tool & Computational Workflow \\
\hline
Author ("creator" for workflow) & Recommended & Minimum \\
\hline
Citation & Recommended & Recommended \\
\hline
Code repository & Optional & \\ 
\hline
Description & Minimum & Recommended \\
\hline
ID & Minimum & Minimum \\
\hline
Input & Optional & Minimum \\
\hline
Name & Minimum & Minimum \\
\hline
License & Recommended & Minimum \\
\hline
Output & Optional & Minimum \\
\hline 
Programming Language & Optional & Minimum \\
\hline
Software version ("version" for workflows) & Recommended & Minimum \\
\hline
Software requirements & & Recommended \\
\hline
\end{tabular}
\end{table}

To standardise the values of the schema, enriching the workflow metadata we use the EDAM ontology \cite{ison_2013} that provides concepts for data types and formats, as well as topics and operations. EDAM concepts are used to semantically annotate inputs and outputs of steps, data types processed by the workflows, and operations performed within the workflows, thereby enhancing both the searchability and interoperability of workflows.

Although developed by the bioscience community, the schema.org profiles have no biology specifics and are agnostic to discipline. EDAM is also including more concepts that are non-bioinformatics -- topics to support ecology, mathematics, language, physics, chemistry, etc. -- and besides can be replaced by other thematic ontologies in the schema.org descriptions.

Alongside the schema and ontology Semantic Artefacts for \MakeUppercase{FAIR} metadata frameworks is a third, \textit{mappings,} for representing crosswalks between schemas and mappings between ontology terms. Mappings are critical for interoperability in an ecosystem of services and a diversity and legacy of standards. Mappings need to be described and managed in their own right \cite{nyberg_2024}. Workflows are software, and recent developments have developed the use of schema.org and JSON-LD Linked Data for metadata for software \cite{europeancommission_2020}, including Research Software APIs and Connectors\footnote{\url{https://faircore4eosc.eu/eosc-core-components/eosc-research-software-apis-and-connectors-rsac}} in the European Open Science Cloud, and, notably, CodeMeta \cite{gruenpeter_2024}. Like our Workflow and Tools Bioschema profiles, CodeMeta adopts schema.org terms from "SoftwareSourceCode" and "SoftwareApplication" but also introduces additional properties that play an important role in software metadata records such as build instructions. 

Other metadata components in our framework inherited from the software community is citation.cff\footnote{\url{https://citation-file-format.github.io/}},  is a plain text file format used to describe software to make it easier for researchers to cite in scholarly publications,  placed in Git repositories for harvesting.

These metadata schema when associated  with workflows and the steps in the workflow, along with similar schema for describing data (e.g., DCAT) feed community scholarship knowledge graphs such as DataCite's PID Graph and openAIRE's Scholarly Knowledge Graph for findability and citation. 

\subsubsection{A Canonical Workflow Description}

Our Collaboratory is agnostic to the many workflow languages. The Common Workflow Language (CWL), provides an interoperable and canonical description for workflows across workflow languages and WMS, the structure for input and output specifications, step-by-step process descriptions, data flow between steps and resource requirements.  Full CWL is executable; Abstract CWL is a variant confined to just the description and does not replace the original language for workflow executions.  The EDAM Ontology annotation is used for strong typing inputs and outputs, within Abstract CWL. An alternative to CWL is the Workflow Description Language (WDL).

\subsubsection{A \MakeUppercase{FAIR} Digital Object Format}

Once a workflow can be described, it then needs to be exchanged across platforms and services within the ecosystem. For supporting findability, accessibility, interoperability and reusability of workflows we need a \MakeUppercase{FAIR} Digital Object format that describes and packages workflows, metadata, associated companion data objects, execution provence logs for reporting, archiving, citation and exchange between services and deposition into repositories.

RO-Crate is a community-developed standardised approach for research output packaging with rich metadata \cite{soilandreyes_2022}, also schema.org and JSON-LD (Linked Data) based \cite{soilandreyes_2024-1}.

\begin{figure}[htbp!]
\centering
\includegraphics[width=.9\textwidth]{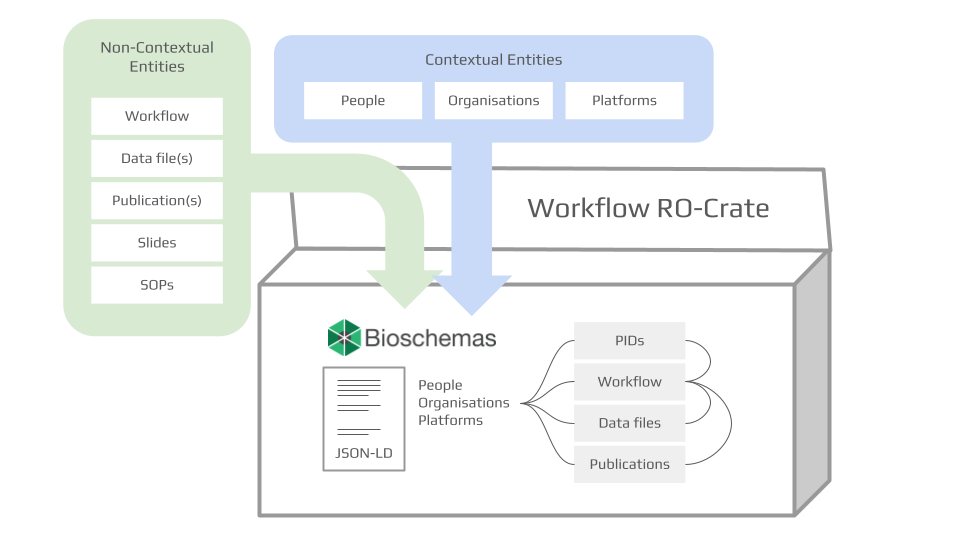}
\caption{\label{fig:2}An RO-Crate for a Workflow with bioschemas, CWL and PIDs, including Contextual Entities. Non-contextual entities can be "attached" (included) or "detached"  (referenced). Contextual Entities refer to PIDs such as An RO-Crate for a Workflow with bioschemas, CWL and PIDs, including Contextual Entities. Non-contextual entities can be "attached" (included) or "detached" (referenced). Contextual Entities refer to PIDs such as ORCID, ROR, and RAiD in Table~\ref{tab:3}. Illustration based on~\cite{SR_elixir_23}.}
\end{figure}

RO-Crate provides the Collaboratory  with the ability to package executable workflows, their components, such as example and test data, abstract CWL, diagrams and their documentation, using the Bioschemas description. An example is illustrated in Figure~\ref{fig:2}. RO-Crates become a currency of exchange between the services of the Workflow Collaboratory \cite{soilandreyes_2024-2,goble_2021}. An RO-Crate profile is a specification or set of guidelines that define how an RO-Crate should be structured and interpreted for a particular use case or domain. It defines:
\begin{compactitem}
    \item Required and optional metadata fields (e.g.,, mandatory author information, dataset descriptions)
    \item Specific ontologies or vocabularies to be used (e.g.,, DCAT, schema.org, domain-specific ontologies)
    \item Constraints on file organization and relationships (e.g.,, linking workflows, datasets, and software in a predefined way)
    \item Validation rules to ensure compliance with community standards
\end{compactitem}

Three RO-Crate profiles have been developed:
\begin{compactitem}
    \item Workflow-RO-Crate\footnote{\url{https://w3id.org/workflowhub/workflow-ro-crate/1.0}} describes and packages the various components required to understand and execute a workflow;
    \item Workflow Testing RO-Crate is used to support the submission of test suites for computational workflows;
    \item Workflow-Run-RO-Crate (WRROC) \cite{leo_2024} is an extended profile that logs and data flow generated by executing a workflow.
\end{compactitem}

The RO-Crates benefit the framework described here by making workflows more portable and reusable, as they are readily exchanged between its services and platforms. While workflow provenance focuses on capturing detailed, technical records of workflow execution, including inputs, outputs, and intermediate steps, transparency aims to provide a broader, more accessible understanding of the entire scientific analysis process. This discrepancy manifests in several ways. Provenance typically offers fine-grained, machine-readable data, whereas transparency requires higher-level, human-readable representations. Approaches are needed that can transform detailed provenance data into more transparent, understandable representations of scientific workflows\footnote{\url{https://iitdbgroup.github.io/ProvenanceWeek2021/t7.html}}.


\subsubsection{Persistent identifiers}

Unique and persistent identifiers are central to FAIR, explicitly appearing in the Findability and Accessibility principles. In this context, where a workflow and each version are assigned an unambiguous, globally unique PIDs. its components are assigned distinct PIDs, and the identifiers are included in their metadata, all of which are retrievable by their identifier. 

Persistence of an identifier supports rediscovery and prevents situations where non-persistent identifiers result in a loss of function. A good example here is the link rot that can affect URLs included in scientific publications. A URL that includes a persistent identifier, such as a DOI, would not lose its function. A persistent source of identity also supports reproducibility in scientific research. There should not be any confusion regarding the identity of a research workflow that has supported a project or publication. Given their importance, persistent identifiers are central to the framework and, as outlined above, are accommodated by the metadata standards from bioschemas.

The metadata in the RO-Crate is associated with Persistent Identifiers for all of these. \textit{Effectively, RO-Crates are metadata objects referencing PIDs.} RO-Crates describe \textit{Data Entities} , which are files within the RO-Crate (called attached) or outside the RO-Crate referenced by a PID (called detached), as \textit{well as Contextual Entities} such as people or places, and conceptual descriptions that primarily exist as metadata, like licenses, which also have globally web-based PIDs. 

\begin{table}[!ht]
\centering
\scriptsize
\begin{threeparttable}
    \caption{PIDs in the Workflow Collaboratory.}
    \label{tab:3}

\begin{tabular}{|p{0.2\textwidth}|p{0.4\textwidth}|p{0.2\textwidth}|}
\hline
Administrative Objects &
\begin{compactitem}
    \item People, e.g., ORCID\tnote{a}
    \item Places, e.g., ROR\tnote{b}
    \item Projects, Funders e.g., RAiD\tnote{c}
\end{compactitem} &
People migrate, places merge and disappear, projects are transitory.\\
\hline

Companion Objects &
\begin{compactitem}
    \item Data repositories for test data or results, e.g., DOIs, Accession Numbers, Handles\tnote{d}
    \item Publications, e.g., DOI
    \item Operating artifacts, e.g., Licenses\tnote{e}
\end{compactitem} &
Limited versioning\\
\hline

Software Objects &
\begin{compactitem}
    \item Workflows, e.g., DOIs when published, PIDs in registries such as WorkflowHub or repositories \cite{price_2024}
    \item Software repositories, e.g., GitHub\tnote{f}, SWHID\tnote{g}
    \item Tools and containers, e.g., PIDs in registries such as bio.tools\tnote{h} and Biocontainers\tnote{i}
\end{compactitem} &
Extensive versioning\\
\hline
\end{tabular}

\begin{tablenotes}
\footnotesize
\item[a] Open Researcher and Contributor IDentification (ORCID), \url{https://orcid.org/}
\item[b] Research Organization Registry (ROR), \url{https://ror.org/}
\item[c] RAiD Identifier Service, \url{https://ardc.edu.au/services/ardc-identifier-services/raid-research-activity-identifier-service/}
\item[d] Example accession number: \url{https://www.ebi.ac.uk/ena/browser/home}
\item[e] SPDX License List, \url{https://spdx.org/licenses/}
\item[f] Example GitHub repo: \url{https://github.com/AustralianBioCommons/Genome-assessment-post-assembly}
\item[g] SoftWare Heritage persistent IDentifiers (SWHIDs), \url{https://docs.softwareheritage.org/devel/swh-model/persistent-identifiers.html}
\item[h] bio.tools registry, \url{https://bio.tools/about}
\item[i] Biocontainers registry, \url{https://biocontainers.pro/}
\end{tablenotes}

\end{threeparttable}
\end{table}

The Persistent Identifiers appearing in our Collaboratory are summarised in Table~\ref{tab:3}.  There are several challenges with PIDs:
\begin{compactitem}
    \item \textbf{Change:} the object that the PID changes; therefore, the PID needs to support versioning.  The \MakeUppercase{FAIR} Principles specify that different versions of the workflow are assigned distinct identifiers (F1.2) and that these identifiers lead to the retrieval of the correct version. Moreover, a workflow is associated with detailed provenance of the workflow (R1.3), which implies its previous versions.
    \item \textbf{Multiplicity:} within a workflow ecosystem, a Workflow (or indeed a dataset or publication) might be assigned multiple PIDS: a DOI, a Git URL, or a registry-specific URL (for example, in a Workflow registry such as WorkflowHub (see section "Workflow Repositories and Registries").
\end{compactitem}

An extensive "PIDs stack" requires policies to define which PIDs an ecosystem will use \cite{vanhorik_2024}. A Workflow and its components are retrievable by their identifiers using a standardised communications protocol (A1) which may include authentication and authorisation steps as \MakeUppercase{FAIR} is not synonymous with Open. A surprising challenge is navigating to the metadata of an object, such as a workflow,  from a landing page to retrieve its machine-actionable metadata and its PID. \MakeUppercase{FAIR} signposting\footnote{\url{https://signposting.org/FAIR/}} uses existing simple web standards (Web Links, HTTP headers, and Linksets) to expose machine-actionable navigation links that indicate downloadable resources, types, and attribution -- particularly for scholarly and institutional repositories that use persistent identifiers like DOIs. 

\subsubsection{Protocols}

It is one thing to be exchangeable, but another to exchange. Here, standards for exchange are important, and the framework makes use of standardised APIs to achieve workflow exchange. The Global Alliance for Genomics and Health (GA4GH) has a basket of relevant APIs for workflows. The GA4GH Tools Registry Service (TRS\footnote{\url{https://www.ga4gh.org/product/tool-registry-service-trs/}}) Application Programming Interface (API) is used to support exchange of both workflows and metadata,  enabling users to search for and retrieve metadata about registered tools, including the tool's name, version, description, author, input and output parameters, and Docker image details. WMS and workflow registries that implement TRS make it possible to both find, retrieve, workflows (F) and launch them (A, R). The GA4GH Workflow Execution Service (WES\footnote{\url{https://www.ga4gh.org/product/workflow-execution-service-wes/}}) API is a protocol for running a workflow in multiple cloud environments, designed to facilitate interoperability between different workflow execution platforms, and hence support workflow reusability across  languages and systems. A RESTful API for submission, monitoring, and management allows workflows to be executed on different platforms without requiring changes to the workflow definition, and with  inputs, outputs, and metadata definitions to ensure reproducibility and traceability. Multiple Workflow Languages are supported including CWL, WDL, and Nextflow. Interoperability across Cloud and HPC Environments enables workflow reusability.

The combination of common metadata, workflow exchange formats, and standardised exchange mechanisms creates several benefits: (i) discoverability through standardised metadata, including identifiers, that makes workflows easier to find and understand, (ii) interoperability through common descriptions that allow workflows to be shared and integrated more easily, and (iii) automation through machine-readable descriptions that enable automated processing and execution of workflows.

\subsection{An Ecosystem of Interoperable Services}
Our ecosystem of interoperable services presented in Figure~\ref{fig:1} makes use of the metadata standards, identifiers, and protocols described in the previous section. Together these support the lifecycle of workflows, from creation to publishing or archiving, described in the next section. Community engagement and support plays a central role here as well, particularly for the different WMS, further discussed in the later stakeholders section. 

\subsubsection{Workflow Management and Execution systems}
At the heart of the ecosystem is a plethora of workflow management systems, from Jupyter Notebooks and Python scripts, to generally used systems (e.g., Nextflow, Snakemake, Galaxy, CWL) and specialist systems (e.g., SCIPION, PyCOMPS etc). In the many hundreds of workflow systems known in the scientific landscape\footnote{\url{https://s.apache.org/existing-workflow-systems}} there is variability in each of these dimensions. Fortunately, the results of recent community efforts have been published which present a standardised terminology for workflow management systems \cite{suter_2025}. Within the EOSC-Life Collaboratory, for example, the Galaxy system has a web-based interface to chain tools together and access data stores that are bundled into a managed  instance, for example Galaxy Europe, configured to use cloud and HPC services through Pulsar, a backend cluster network service. Nextflow has a dedicated client application for authoring and management, and a Seqera backend execution platform\footnote{\url{https://seqera.io/nextflow/}} supported by all major commercial cloud platforms, including Microsoft Azure, AWS, and GCP. WDL uses Terra as a backend. Snakemake, CWL, and more specialist workflow systems like SCIPION expect their authors to install the packages and edit their scripts using conventional editors as well as set up their own servers, similar to scripting platforms. In an effort to support pan-platform workflow reuse, the community has developed back-end execution services such as Sapporo and WfExS that execute workflows in different languages through a common interfaces using the GA4GH Workflow Execution Service\footnote{\url{https://www.ga4gh.org/product/workflow-execution-service-wes/}} (WES API) and Workflow-RO-Crates.

Most systems use packaging and containers. On our Collaboratory, tools developed and managed as part of BioConda, a community-driven Conda channel that packages over 9000 bioinformatics software, are semi-automatically converted to Docker containers \cite{bray_2023}. Containers provide a means to package entire programming environments along with their operating systems. Container platforms like Kubernetes\footnote{\url{https://kubernetes.io/}} can be used to orchestrate applications across multi-cloud and hybrid-cloud environments, enabling efficient management of resources across on-premise private clouds (e.g., OpenStack\footnote{\url{https://www.openstack.org/}}) as well as those of the hyperscalers like Amazon Web Services\footnote{\url{https://aws.amazon.com/}}. Tools and containers, such as Docker and Apptainer form part of the workflow build process, as steps in the workflow. In many cases importing these tools and containers is enabled by APIs. Beyond containerization, Infrastructure as Code (IaC) tools such as Terraform and Ansible can play a crucial role in automating the provisioning of compute resources, managing software dependencies, and ensuring consistent configuration across environments. Researchers could define, share, version, and reuse their workflow execution infrastructure e.g., by integrating Terraform and Ansible configurations into GitHub or workflow registries such as WorklowHub as we will see next, further advancing the usability and reusability of computational workflows. In addition, tools like Caper\footnote{\url{https://www.encodeproject.org/software/caper/}} automate the creation of necessary configurations, reducing the manual effort required to set up workflows in different environments.

\subsubsection{Workflow Repositories and Registries}

Workflows need to be developed, managed and saved. Communities, as might be expected, typically use Git (GitLab or GitHub) for shared workflow development with version control. Some communities manage high quality curated collections, such as Galaxy's Intergalactic Workflow Commission\footnote{\url{https://github.com/galaxyproject/iwc}}, Nextflow's nf-core\footnote{\url{https://nf-co.re/}} \cite{ewels_2020}, BioWDL\footnote{\url{https://biowdl.github.io/}} for WDL and KNIME Hub\footnote{\url{https://www.knime.com/knime-hub}} for KNIME\footnote{\url{https://www.knime.com/}}; others have a more laissez-faire attitude such as CWL, which has a CWLViewer\footnote{\url{https://view.commonwl.org/}} \cite{robinson_2017} for browsing multiple GitHub repositories and the Snakemake workflow catalog\footnote{\url{https://snakemake.github.io/snakemake-workflow-catalog/docs/workflows/snakemake-workflows/rna-seq-star-deseq2.html}}.

Repositories tend to be language- or community- or project-focused, or all three. Registries such as Dockstore \cite{yuen_2021}, Methods Hub\footnote{\url{https://www.nfdi4datascience.de/services/gesismethodshub/}}, and WorkflowHub \cite{gustafsson_2025} provide a one-stop shop for workflows to showcase and share beyond their repositories. The WorkflowHub workflow registry:
\begin{compactitem}
    \item is open to any workflow language or discipline: to date the registry has workflows from over 20 different languages and platforms, including KNIME, Jupyter Notebooks, AutoSubmit and COMPSs;
    \item catalogues workflows with rich metadata while they remain in situ in their home Git repositories or execution environments while managing automated registration and versioning processes;
    \item encourages curated collections that can cross language, project and community borders;
    \item supports Communities of Practice (Teams and Spaces) to organise, curate and share workflows to reflect their project and community origins. As well as open access, the Hub supports private "enclave" sharing within communities as well as public sharing (\MakeUppercase{FAIR} is not the same as Open). The Life Science Login (LS-Login\footnote{\url{https://lifescience-ri.eu/ls-login.html}}) service enables researchers to use their home organisation credentials or community or other identities to sign in and access private workflows, and WorkflowHub to control and manage access rights.
    \item supports workflow publishing with minted DOIs for workflow entries and integration with the scholarly communications infrastructure.
\end{compactitem}

\begin{figure}[!ht!]
\centering
\includegraphics[width=.9\textwidth]{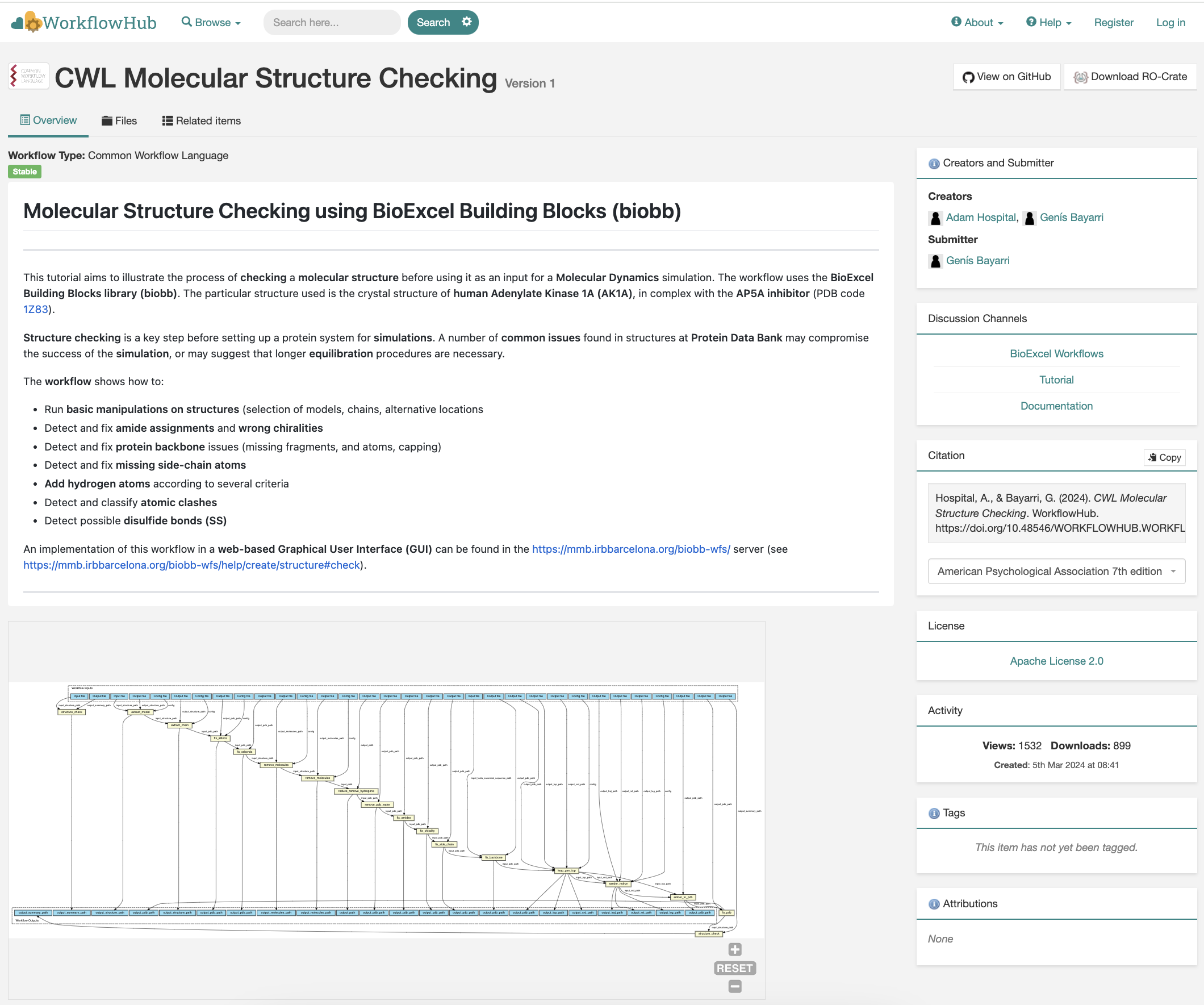}
\caption{\label{fig:3}WorkflowHub entry \cite{hospital_2024}.}
\end{figure}

The WorkflowHub uses our rich metadata framework for findability with inter-registration integration with sister registries for containers (BioContainers \cite{daVeigaLeprevost_2017}), and tools (bio.tools \cite{ison_2019}) using GA4GH APIs and shared PIDs. An example WorkflowHub entry is shown in Figure~\ref{fig:3}. Tools and Workflows are marked up with metadata standards from Bioschemas, encouraging the enrichment of metadata by the workflow and tool developers, including licensing and provenance metadata. Abstract CWL is recommended as a canonical representation of the workflow, and the Hub imports and exports the exchange standard Workflow-RO-Crates, facilitating automatic builds for each of its entries, regardless of whether the management system natively supports it. 

For accessibility and reusability, WorkflowHub implements the TRS API for a standardised communication protocol to launch workflow executions for those execution platforms that implement it (e.g.,\ Galaxy) without requiring bespoke integrations. Conversely WorkflowHub can be searched from within the execution platform using the same API.

WorkflowHub has its own PID and supports the minting of associated DataCite DOIs for public, open workflows. In keeping with the Accessibility principles, all workflows are resolvable by the PID but not always accessible without authorisation and authentication (using LS-Login). WorkflowHub entries are annotated with tool PIDs to link through to bio.tools entries, and bio.tools entries link to WorkflowHub PIDs that use a given tool. WorkflowHub entries are associated by PIDs with other companion objects such as test data, publications and other workflows, ORCID ids for people, ROR ids for institutions etc. WorkflowHub is an early adopter of \MakeUppercase{FAIR} signposting, using the patterns Bibliographic Metadata and Publication Boundary to signpost the Workflow RO-Crate for automated findability of the machine-actionable metadata held by the registry.

One of the most important PIDs managed by WorkflowHub is the Git iD entry for the workflow repository.  Although the registry will accept file uploads, its intention is to index the workflows in their native repositories, and should those repositories disappear, the workflow metadata will still be accessible on WorkflowHub. Integration with Git repositories requires an automated registration process, including harvesting local citation.cff files to expose workflow author credit for citation, with life cycle support around versioning with Git support, and distinct PIDs for different versions of the workflow.

\subsubsection{Testing and benchmarking}
As workflows are software and software is in a constant state of development and maintenance, our ecosystem needs services that focus on usability as well as reusability of tools and workflows. Containers, dependency management and \MakeUppercase{FAIR} unit testing of workflow components also contribute to workflow (re)usability. LifeMonitor\footnote{\url{https://lifemonitor.eu/}} monitors and triggers automated workflow tests and automated checks on metadata and adherence to best practices on the workflow's source code Git repository. OpenEBench\footnote{\url{https://openebench.bsc.es/}} benchmarks tools, and monitors software quality as well as scientific benchmarking to help determine the precision, recall and other metrics of bioinformatics resources in unbiased scenarios. The OpenEBench observatory provides automated monitoring of \MakeUppercase{FAIR} for research software metrics and indicators. The metadata framework, TRS APIs and specialised Workflow-Testing-RO-Crate and WorkflowRun-RO-Crates are used as the currency for exchanging workflow metadata between WorkflowHub, these testing services and the execution platforms tested.

\subsubsection{Research Information System services, publishers and repositories}
This ecosystem of services plugs into the wider services of scholarly communication to support publication, citation and knowledge discovery. WorkflowHub's use of ORCID and DataCite DOIs supports workflow author credit and citation and contributes to the DataCite PID Graph and OpenAIRE Research Graph. The WorkflowHub is used by publishers as a recommended resource for publishing workflows associated with publications, with access to the cited version and new versions. The RO-Crate packaging means that every workflow in WorkflowHub can be deposited in other long term repositories like Zenodo or DataVerse, or even Software Heritage. The workflow or its execution software might not exist, but the description still will, giving a long term commitment to accessibility. The \MakeUppercase{FAIR} Digital Object approach -- using RO-Crate and a metadata framework and PIDS -- means we can package links to data, cleanly reference the software used, and begin to build citation metrics.

\subsubsection{Onboarding \MakeUppercase{FAIR} Services and Standards}
WorkflowHub is a convergence point for workflows in the Collaboratory, supporting the metadata framework, and the ability for a user of the ecosystem to engage with it in a meaningful way \cite{gustafsson_2025}. WorkflowHub automated build of Workflow RO-Crate allows a user to begin exchanging their workflow with other infrastructures, including compute via WfExS, and testing via LifeMonitor.

Although the Hub takes all forms of workflows and annotates them with as much metadata as possible, the real power of the \MakeUppercase{FAIR} ecosystem is when the Workflow Management Systems step up to support the framework. Extracting and generating metadata from community-accepted standard formats is a key aspect of the practical use of the framework and the Collaboratory. Our most popular workflow management systems (Galaxy, Snakemake, CWL and Nextflow) have onboarded our metadata framework with:
\begin{compactitem}
    \item Native support for the metadata framework through RO-Crate enrichment of RO-Crates as a workflow evolves and crosswalks to the required Bioschema attributes for supporting registration in WorkflowHub;
    \item Native support for Workflow-Run-RO-Crates for provenance and links to data products;
    \item Native support for automated Git bot registration and version updates from curated repositories Galaxy IWC and nf-core;
    \item Implementation of TRS API so workflows can be found on WorkflowHub from within a Galaxy instance  and workflows found on WorkflowHub can be directly accessed, launched and reused from the registry entry, in Galaxy, Sapporo or WfExS.
\end{compactitem}

In some cases, notably Galaxy, the integration is very rich and goes further to include:
\begin{compactitem}
    \item Bio.tools PIDs in the native workflow language description;
    \item Support for Abstract CWL generated from the native language description;
    \item Data products exported as RO-Crates.
\end{compactitem}

One of the challenges of our ecosystem, as foreshadowed earlier, is PID multiplicity. A workflow will be assigned a PID registration entry by WorkflowHub, minted a DOI registration in WorkflowHub, and a Git URL in its repository and, perhaps, an RO-Crate PID in Zenodo, assigned its own DOI and a Software Heritage ID (SWHID). We choose to use the WorkflowHub PID as the authority, and seek to gather the PIDs and their versions, in the WorkflowHub record.

\subsection{Supporting the \MakeUppercase{FAIR} Workflow Lifecycle}
Our workflow Collaboratory focuses on convergence points for workflows, and exchange between a constellation of services and platforms surrounding the lifecycle of the workflow. Workflows cannot be separated from their data;  a workflow's lifecycle cannot be separated from the lifecycle of its data. When considering how to support the \MakeUppercase{FAIR} workflow lifecycle, it is important to consider what tools and services are necessary to make both a workflow \textit{and its data products} \MakeUppercase{FAIR} by design.

\begin{figure}[!ht]
\centering
\includegraphics[width=.9\textwidth]{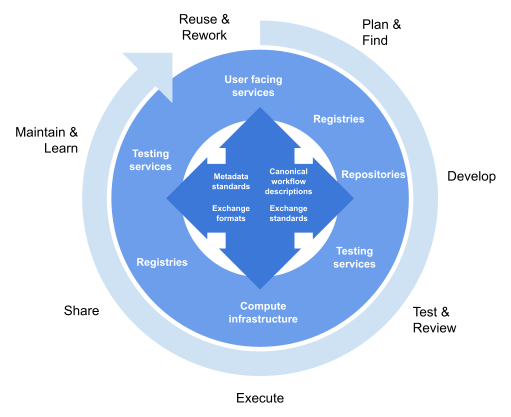}
\caption{\label{fig:4}The Workflow Lifecycle: \MakeUppercase{FAIR} by Design. 
Life cycle adapted from \cite{gustafsson_2025}.}
\end{figure}

The \MakeUppercase{FAIR} workflow lifecycle resembles the \MakeUppercase{FAIR} software lifecycle, except that there are additional considerations to incorporate the \MakeUppercase{FAIR} data lifecycle, too. Figure~\ref{fig:4} presents a characterization of the workflow lifecycle. This lifecycle emphasises the processes and considerations along with the respective services necessary to ensure FAIRness throughout the workflow's existence, from initial design to long-term maintenance and reuse.

\textbf{\textit{Plan and Find:}} A workflow developer may try to find relevant existing workflows or their components on the basis that the majority of workflows or scripts are variants, and workflows are compositional and made up of building blocks. WorkflowHub, BioContainers, and Bio.tools are our cross-integrated registries for this purpose, and support rich metadata as well as indicators of who built the workflows and for what purpose. In practice metadata quality depends on the support of the workflow management system and the diligence of the workflow curator. 

\textbf{\textit{Workflow development:}} A workflow or its component "can be understood, modified, built upon or incorporated into other workflows" (I3), and are largely developed iteratively in version-controlled Git repositories, which WorkflowHub keeps track of.  Interoperability and reusability place obligations on workflow developers. Workflows written in common languages (e.g., CWL, WDL) and specified in ways that allow the components to read, write, and exchange data in standard ways (e.g., RO-Crate) supports interoperability, but this depends on support offered by the Workflow Managers. In practice developers need to ensure that tools and datasets have clean I/O programmatic interfaces, no usage restrictions, use of community data standards and identifiers, and that they are simple to install and designed for portability. Workflow developers can be both data-FAIR, by using and making identifiers, licensing data outputs, tracking data provenance and so on, and workflow-\MakeUppercase{FAIR} by managing versions, providing test data, and sharing libraries of composable and reusable workflow "blocks" to be used as subworkflows. Reusing workflows or components requires clear and compatible licensing indicated in the registries; tools need to be workflow ready \cite{brack_2022}. Thus \MakeUppercase{FAIR} applies "all the way down" for the whole workflow and each of its components. Communities such as BioBB \cite{andrio_2019}, MGnify \cite{richardson_2023}, and nf-core \cite{ewels_2020} are working on reviewing, validating, and certifying libraries of reusable, composable canonical workflows designed to be \MakeUppercase{FAIR} unit tested and to generate \MakeUppercase{FAIR} data. These are registered  in curated collections in WorkflowHub. 

\textbf{\textit{Test \& Review:}} The workflow is executed and monitored. Human in the Loop workflows may involve user intervention, but tests are also performed by tester workflows. The EOSC-Life Collaboratory uses OpenEBench and LifeMonitor to monitor tests and review workflows, gathering and analyzing provenance in RO-Crates to determine if the workflow behaves as intended, and continues to do so, reporting to WorkflowHub. Principles I1, I4, and R1.3  describe \MakeUppercase{FAIR} practices for provenance which help directly with testing and reviewing, and execution.

\textbf{\textit{Execute:}} Workflow execution results in \MakeUppercase{FAIR} data products and provenance data describing specific workflow executions, in WorkflowRun RO-Crates, as well as collecting execution related information such as resource usage needed for improving quality of service, optimizing resource allocation, and identifying bottlenecks. This metadata facilitates reuse, for example porting a workflow across computational resources with different capacities or processing significantly larger datasets than previously. OpenTelemetry\footnote{\url{https://opentelemetry.io/}} \cite{blanco_2023}, for instance, facilitates portable telemetry by providing standardised APIs, SDKs, and tools for collecting and analyzing execution data.

\textbf{\textit{Share:}} The driver for \MakeUppercase{FAIR} is to share: (i) the workflow components and the workflow written in its language with instructions for configuration and execution with links to test data, and  ii) the results of the workflow run with links to its provenance records and data products. Documentation includes metadata and annotations and are aimed for third-party users to reuse or repurpose the workflow or understand and gain insights from the workflow execution. In our Collaboratory the workflow components and workflow designs are developed in Gits and registered in WorkflowHub; the workflow run records are deposited in data repositories as Workflow-Run-RO-Crates, and occasionally registered in WorkflowHub to be published to support publications. WorkflowHub supports publishing  (by minting a DOI, which means that the entry is fixed) and restricted sharing (by providing elaborate sharing permissions for Teams that restrict access (A1.2), and curators control the visibility). 

\textbf{\textit{Maintain \& Learn:}} This phase addresses the long-term relevance and effectiveness of the workflow, and includes updating workflow versions to maintain compatibility while refining components, adapting the workflow to new datasets, methods, or computational environments, and ensuring its interoperability across different workflow systems. This phase is community-driven and generally requires input from several users/developers. Active community engagement and continuous improvements ensure the long-term sustainability and FAIRness of the workflow.

\textbf{\textit{Reuse \& Rework:}} Back to development, but this time emphasising the importance of citing workflows, crediting their developers and maintaining the provenance of a \textit{workflow design} (R1.3) rather than the workflow execution, and references to other workflows (R2) it uses or was derived from.  WorkflowHub includes "attributions" that a Workflow has; that is workflows that it reuses as a sub-workflow or a workflow it was forked or cloned from and produces a KnowledgeGraph to track workflow and other objects interrelations. In practice attributions are rarely made even when they should or could be and are hard to identify without support by the workflow management systems. CWL has support for explicit subworkflows identification and as does a Galaxy (partially) when a workflow is loaded from WorkflowHub. 

\subsection{\MakeUppercase{FAIR} Actors: Workflows Need a Village}
Just as it takes a village to manage and share data \cite{borgman_2022}, \MakeUppercase{FAIR} workflows need a village, too. There are different kinds of stakeholders (human or machine) in a computational ecosystem who have different needs and fulfill the different roles illustrated in Figure~\ref{fig:1}.

\textbf{\textit{Tool developers}} produce the artifacts that are essential to any functioning workflow ecosystem, in particular data and software tools. Tool developers have produced thousands of software packages and API-accessible tools  to aid biomedical research but still need to work to make their tools workflow-friendly \cite{brack_2022}, with clean interfaces and non-restricting usage conditions.

\textbf{\textit{Workflow developers}} interact with the entire represented ecosystem through their connections to data, software tools, computational workflows, and infrastructure. They are the creators and maintainers of workflows as digital objects, and they leverage elements of the ecosystem to address their core business of building, testing, and optimizing workflows for specific computational workloads (i.e. different data analyses processes), for different data types, and for execution on a number of diverse computational infrastructures. They are also responsible for designing workflows for interoperability and reuse, and registering them to enable findability and accessibility. Communities of workflow developers build well curated and canonical workflows that we can address directly in order to be able to improve their practices. 

\textbf{\textit{System administrators}} are intimately connected to the underlying computational infrastructure in an ecosystem. Users' needs inform the design of infrastructure \cite{vazhkudai_2018} (e.g., what workflows they intend to execute), just as the suitability of available infrastructure informs the design choices made in workflows by their developers. An execution infrastructure is often selected based on its suitability for a given WMS, the computational resource requirements of the processes defined by the workflow, and the availability of systems that ease the deployment of a workflow to that infrastructure. For example, Galaxy workflows require an operational Galaxy instance, with the correct tools installed, and the underlying compute resource to execute the workflow. The responsibilities of a system administrator frequently include managing both software and hardware. System administrators use workflows to streamline processes and automate these orchestrations, and testing frameworks like LifeMonitor help to support the sustainability and reusability of published computational workflows.

\textbf{\textit{Curators}} streamline the ecosystem for all users: they ensure that all data and metadata – whether in data repositories, software repositories, container registries, workflow registries, or otherwise – are organised and maintained in ways that promote open, efficient, and sustainable research practices. A curator ensures that these resources are properly documented, tagged, and described with relevant metadata, which facilitates their discoverability and usability across different systems and disciplines. By enforcing the \MakeUppercase{FAIR} Principles, curators enable other users to locate, retrieve, and repurpose data and workflows without unnecessary barriers. Curators play a critical role in establishing and maintaining interoperability among different resources within the ecosystem. As digital artifacts come from various sources, curators must ensure that they adhere to common standards and formats that allow for seamless integration and communication across tools, repositories, and platforms. This helps to preserve the longevity and reusability of digital artifacts over time, ensuring that they remain useful and usable as technologies evolve. The curator also oversees the processes for updates and version control, ensuring that the latest versions of datasets, code, and workflows are properly synchronised while also ensuring that they remain compliant with evolving \MakeUppercase{FAIR} standards. In essence, curators safeguard the ecosystem's integrity and functionality, empowering other users to build on each other's work effectively and efficiently.

\textbf{\textit{End users}} are the bioinformaticians and researchers. We want to encourage people to use well-documented, FAIR-enabling, and \MakeUppercase{FAIR} workflows and to credit the workflow builders because this is a non-trivial and expensive activity. End users might download an available dataset using a web browser-based portal for manipulation and analysis workflows or run a workflow on their own infrastructure and their own datasets. The end users range across a spectrum from expert workflow developers looking to reuse or repurpose workflows on their own infrastructure to no-code analysers finding and launching a carefully configured workflow with limited inputs for datasets and parameter settings.  In a fully developed ecosystem, these latter end users would be insulated from many of the details which are currently required knowledge to achieve their goals, such as exactly what graphics processing unit (GPU) is required. An ecosystem like the EOSC-Life Collaboratory seeks to create an open, digital, and collaborative space for European life science research by making biological and medical research data and other Digital Objects \MakeUppercase{FAIR} across disciplines, allowing scientists to easily access, integrate, and analyze life science data from various sources for research purposes, ultimately driving innovation in the field by enabling wider data sharing and collaboration. Where the Collaboratory excels is in additionally promoting the FAIRness of the tools/software, and workflows that will be used to access, integrate, and analyze the data. Aspiring for FAIRness "all the way down" leads to a \MakeUppercase{FAIR} ecosystem and a number of advantages for both human and machine users of the future.

Community activism, led by the platforms and registries coming together in a community group like the Workflows Community Initiative\footnote{\url{https://workflows.community/}}, is needed to define principles, policies, and best practices for \MakeUppercase{FAIR} workflows, standardise metadata representation and collection processes, and work on the standards and communities for building sustainability and policy.

\section{Challenges of a \MakeUppercase{FAIR} Workflow Ecosystem}

Exposing scientific workflow metadata so that they can be found, understood, and re-used by others (i.e. made FAIR) can be extremely challenging due to the sheer number of WMS, the heterogeneity of data, software, and formats, the lack of interoperability, the different paradigms workflows can be predicated on, as well as complex workflow lifecycles that encompass specification, execution, data products, and metrics \cite{pouchard_2022}. As a result, there is still a need to effectively implement the \MakeUppercase{FAIR} Principles, while taking into account this complexity \cite{wilkinson_2022}. In particular, we need to be able to automate the FAIRness in workflows, check that development was FAIR, and that FAIRness extends to the data flowing through workflows. It is important for workflow platforms to enable \MakeUppercase{FAIR} outputs, including correct citation of input data and the software used.

\subsection{Standardization}

Workflow FAIRification is still in its early days and suffers from a lack of standardisation across disciplines. Even if workflow components from distinct domains are findable and accessible, because they are treated either as \MakeUppercase{FAIR} data or \MakeUppercase{FAIR} software, both interoperability and reusability require a minimum degree of standardization. The ideal situation would ensure that newly developed workflow components align to the same standard as existing components. In reality, building a \MakeUppercase{FAIR} computational workflow is complicated by the diversity in both the design and maturity of the components that are being added. Many datasets and software do not comply with all \MakeUppercase{FAIR} principles, and there is a spectrum of \MakeUppercase{FAIR} compliance. As a result, it may be necessary to include additional modules that support chaining and interoperability for non-\MakeUppercase{FAIR} components. For example, APIs that manage data format translations to streamline data flow between workflow components. However, requiring that all components are \MakeUppercase{FAIR} may serve to establish an unnecessarily high bar for workflow FAIRification, particularly in domains where an ecosystem of \MakeUppercase{FAIR} components is not yet well established.

Maintaining and adopting standards is an ongoing challenge in the rapidly changing landscape of scientific computing, requiring regular updates to existing standards (e.g., CWL) and the incorporation of new technologies. For example, containerization technologies (i.e. Docker, Apptainer) drove the adaptation of workflow standards to include container specifications, cloud computing drove standards for describing cloud-based workflow execution environments, and ontologies like EDAM are continuously updated to reflect new topics, operations, and data formats. Collaborative efforts (e.g., Research Data Alliance, RDA), play a crucial role in ensuring that standards evolve in lock step with technological advancements and community requirements.

FAIRification of a workflow specification based on DASK.distributed is an example of creating a \MakeUppercase{FAIR} ecosystem by adding identifiers, capturing data and provenance for both different layers of the software stack and execution environment, and finally reconciling these data into a common format where variables can be correlated for analysis \cite{gueroudji_2024}. When combined with composable, interoperable tools, this example illustrates that it is possible to propagate requirements for findability and interoperability.

\subsection{Workflow Portability}

Workflow portability, derived from software portability, refers to workflows' ability to run across different platforms. However, workflows face challenges due to their composable nature -- they are only as portable as their least portable components. Portability for components such as software is often hindered by dependencies on specific environments and/or specialised hardware, both of which can be difficult to replicate. When workflows rely on complex software stacks or custom configurations, they may fail in new environments due to variations in system setups or library versions (e.g., legacy operating systems, institutional security measures, batch queueing systems). The \MakeUppercase{FAIR} Principles -- especially Findability and Interoperability – can help by promoting clear documentation of dependencies and configurations, making it easier to reconstruct environments. 

At this point, there is always someone in the crowd who believes that containers are a perfect solution because of their ability to encapsulate workflows alongside their environments; they are mistaken. Containers are built for specific hardware, and they fail when hardware assumptions are not met, just as compiled executables do. Most workflows are not hardware-dependent, but software components in HPC workflows frequently rely on exotic hardware (e.g., GPUs, NVMe, SSDs) and implicit assumptions (e.g., number of cores, amount of memory, network bandwidth) that do not translate well to even to other supercomputers, much less smaller machines like laptops or workstations. Frequently, home directories mount differently on login nodes than on compute nodes. Distributed file systems present the familiar POSIX interface but require different assumptions. Workflows may depend on persistent services such as databases that need a place to run, and HPC centers vary in how they provide that infrastructure. The \MakeUppercase{FAIR} Principles help in these cases by encouraging rich metadata and documentation of hardware needs and dependencies. The "FAIR-aware" workflows of the future will be able to adapt dynamically based on information about software environment, hardware availability, and even remaining user quotas, increasing long-term sustainability and adaptability \cite{wilkinson_2025-2}.

\subsection{Reproducibility \& Transparency}

Sometimes mistakenly thought to be the "R" in FAIR, reproducibility typically refers to the results of executing a workflow, rather than the workflow itself. The \MakeUppercase{FAIR} Principles are essential, however, for improving the reproducibility and transparency of workflows, particularly in ensuring that the scientific experiments and data analyses they encode are accessible and verifiable. Transparency goes beyond making the workflow specification itself FAIR; it requires that all key artifacts -- such as datasets, results, and provenance information -- are also FAIR-compliant, which enables better traceability, verifiability, and reuse of computational experiments. While there have been recent advancements in frameworks, specifications, and tools to support this, there is still a need for greater emphasis on making workflows more reproducible and transparent. Current documentation often provides only a high-level overview, which is insufficient for capturing the complexity and dependencies within workflows. To address this, solutions that enable finer-grained workflow annotations -- readable by both humans and machines -- are needed. Detailed annotations would allow for automated processing, greater interoperability, and enhanced understanding, aligning with the broader goal of making workflows more reproducible, transparent, and reusable in scientific research.

\subsection{Quality}

The challenges posed by workflow quality in \MakeUppercase{FAIR} workflow ecosystems include the need for community-agreed metrics to evaluate whether workflows, repositories, or dataflows are truly \MakeUppercase{FAIR} \cite{ferreiradasilva_2024}. Workflow shimming \cite{mohan_2014}, which resolves incompatibilities between tasks, can reduce adherence to \MakeUppercase{FAIR} principles by introducing time-consuming transformations and hindering reusability and interoperability, especially since shims are often not portable across workflow management systems. Solutions like more expressive type systems \cite{kashlev_2014} and data assembly lines \cite{zinn_2009} can help address these issues, but broader adoption remains essential. Initiatives like the Galaxy Project and WorkflowHub are helping promote \MakeUppercase{FAIR} workflows by making them more accessible and reusable, while funding agencies and journals increasingly mandate \MakeUppercase{FAIR} data practices, indirectly promoting the use of \MakeUppercase{FAIR} workflows. However, maintaining standards amidst evolving technologies, such as containerization and cloud computing, presents an ongoing challenge. To aid adoption, tools like the CWL Viewer enable visualizations of workflows; RO-Crate packages research data, its metadata, and its related workflow descriptions together; and validation tools like cwltool help ensure adherence to the CWL standard. As standards and tools evolve, they play a key role in lowering the barriers to creating, sharing, and improving the quality of \MakeUppercase{FAIR} workflows across the scientific community.

\section*{Acknowledgments}

This work was primarily supported by EOSC-Life (WP2 Tools Collaboratory) (H2020 INFRAEOSC, 824087), EuroScienceGateway (HORIZON-INFRA-2021-EOSC-01-04 1010-57388, UKRI 10038963), and BY-COVID (HORIZON-INFRA-2021-EMERGENCY-01 10-1046203), and partially supported by Biodiversity Genomics for Europe (HORIZON-CL6-2021-BIODIV, UKRI 10040409), BioDT (HORIZON-INFRA-2021-TECH-01-01 101057-437, UKRI 10038930), EOSC4Cancer (HORIZON-INFRAEOSC  101058427, UKRI 1003-8961), and PREP-IBISBA (H2020-EU.1.4.1.1., 871118). This research used resources of the Oak Ridge Leadership Computing Facility at the Oak Ridge National Laboratory, which is supported by the Office of Science of the U.S. Department of Energy under Contract No. DE-AC05-00OR22-725. This manuscript has been authored by UT-Battelle, LLC, under contract DE-AC05-00OR22725 with the US Department of Energy (DOE). This written work is authored by an employee of NTESS. The employee, not NTESS, owns the right, title and interest in and to the written work and is responsible for its contents. Any subjective views or opinions that might be expressed in the written work do not necessarily represent the views of the U.S. Government. The US government retains and the publisher, by accepting the article for publication, acknowledges that the US government retains a nonexclusive, paid-up, irrevocable, worldwide license to publish or reproduce the published form of this manuscript, or allow others to do so, for US government purposes. DOE will provide public access to these results of federally sponsored research in accordance with the DOE Public Access Plan (\url{https://www.energy.gov/doe-public-access-plan}). This work is supported by Australian BioCommons which is enabled by NCRIS via Bioplatforms Australia funding. The authors would like to acknowledge the many contributions of the users of the EOSC-Life Collaboratory.

\printbibliography{chapters/50/fair-workflows/references}

\end{document}